\documentclass[12pt]{article}
 \usepackage{epsfig}
\newcommand{\be}{\begin{equation}}
\newcommand{\ee}{\end{equation}}
\newcommand{\beq}{\begin{equation}}
\newcommand{\eeq}{\end{equation}}
\newcommand{\bea}{\begin{eqnarray}}
\newcommand{\eea}{\end{eqnarray}}

\begin{document}
\baselineskip=15.5pt
\pagestyle{plain}
\setcounter{page}{1}
%--------+---------+---------+---------+---------+---------+---------+
%Body

% Ofer's definitions

\def\del{{\partial}}
\def\vev#1{\left\langle #1 \right\rangle}
\def\cn{{\cal N}}
\def\co{{\cal O}}
\newfont{\Bbb}{msbm10 scaled 1200}     %instead of eusb10
\newcommand{\mathbb}[1]{\mbox{\Bbb #1}}
\def\IC{{\mathbb C}}
\def\IR{{\mathbb R}}
\def\IZ{{\mathbb Z}}
\def\RP{{\bf RP}}
\def\CP{{\bf CP}}
\def\Poincare{{Poincar\'e }}
\def\tr{{\rm tr}}
\def\tp{{\tilde \Phi}}

\def\TL{\hfil$\displaystyle{##}$}
\def\TR{$\displaystyle{{}##}$\hfil}
\def\TC{\hfil$\displaystyle{##}$\hfil}
\def\TT{\hbox{##}}
\def\HLINE{\noalign{\vskip1\jot}\hline\noalign{\vskip1\jot}}
\def\seqalign#1#2{\vcenter{\openup1\jot
  \halign{\strut #1\cr #2 \cr}}}
\def\lbldef#1#2{\expandafter\gdef\csname #1\endcsname {#2}}
\def\eqn#1#2{\lbldef{#1}{(\ref{#1})}%
\begin{equation} #2 \label{#1} \end{equation}}
\def\eqalign#1{\vcenter{\openup1\jot
    \halign{\strut\span\TL & \span\TR\cr #1 \cr
   }}}
\def\eno#1{(\ref{#1})}
\def\href#1#2{#2}
\def\half{{1 \over 2}}

%--------+---------+---------+---------+---------+---------+---------+
%Hirosi's macros:
\def\ads{{\it AdS}}
\def\adsp{{\it AdS}$_{p+2}$}
\def\cft{{\it CFT}}

\newcommand{\ber}{\begin{eqnarray}}
\newcommand{\eer}{\end{eqnarray}}

\newcommand{\beqar}{\begin{eqnarray}}
\newcommand{\cN}{{\cal N}}
\newcommand{\cO}{{\cal O}}
\newcommand{\cA}{{\cal A}}
\newcommand{\cT}{{\cal T}}
\newcommand{\cF}{{\cal F}}
\newcommand{\cC}{{\cal C}}
\newcommand{\cR}{{\cal R}}
\newcommand{\cW}{{\cal W}}
\newcommand{\eeqar}{\end{eqnarray}}
\newcommand{\tht}{\thteta}
\newcommand{\lm}{\lambda}\newcommand{\Lm}{\Lambda}
\newcommand{\eps}{\epsilon}

%--------+---------+---------+---------+---------+---------+---------+

\newcommand{\nonu}{\nonumber}
\newcommand{\oh}{\displaystyle{\frac{1}{2}}}
\newcommand{\dsl}
  {\kern.06em\hbox{\raise.15ex\hbox{$/$}\kern-.56em\hbox{$\partial$}}}
\newcommand{\id}{i\!\!\not\!\partial}
\newcommand{\as}{\not\!\! A}
\newcommand{\ps}{\not\! p}
\newcommand{\ks}{\not\! k}
\newcommand{\D}{{\cal{D}}}
\newcommand{\dv}{d^2x}
\newcommand{\Z}{{\cal Z}}
\newcommand{\N}{{\cal N}}
\newcommand{\Dsl}{\not\!\! D}
\newcommand{\Bsl}{\not\!\! B}
\newcommand{\Psl}{\not\!\! P}
\newcommand{\eeqarr}{\end{eqnarray}}
\newcommand{\ZZ}{{\rm \kern 0.275em Z \kern -0.92em Z}\;}
%--------------------------------Alfonso definitions%%%%%%%%%%%%%

% DEFINITIONS

\def\del{{\delta^{\hbox{\sevenrm B}}}} \def\ex{{\hbox{\rm e}}}
\def\azb{A_{\bar z}} \def\az{A_z} \def\bzb{B_{\bar z}} \def\bz{B_z}
\def\czb{C_{\bar z}} \def\cz{C_z} \def\dzb{D_{\bar z}} \def\dz{D_z}
\def\im{{\hbox{\rm Im}}} \def\mod{{\hbox{\rm mod}}} \def\tr{{\hbox{\rm Tr}}}
\def\ch{{\hbox{\rm ch}}} \def\imp{{\hbox{\sevenrm Im}}}
\def\trp{{\hbox{\sevenrm Tr}}} \def\vol{{\hbox{\rm Vol}}}
\def\rl{\Lambda_{\hbox{\sevenrm R}}} \def\wl{\Lambda_{\hbox{\sevenrm W}}}
\def\fc{{\cal F}_{k+\cox}} \def\vev{vacuum expectation value}
\def\nodiv{\mid{\hbox{\hskip-7.8pt/}}}
\def\ie{{\em i.e.}}
\def\ie{\hbox{\it i.e.}}

\def\CC{{\mathchoice
{\rm C\mkern-8mu\vrule height1.45ex depth-.05ex
width.05em\mkern9mu\kern-.05em}
{\rm C\mkern-8mu\vrule height1.45ex depth-.05ex
width.05em\mkern9mu\kern-.05em}
{\rm C\mkern-8mu\vrule height1ex depth-.07ex
width.035em\mkern9mu\kern-.035em}
{\rm C\mkern-8mu\vrule height.65ex depth-.1ex
width.025em\mkern8mu\kern-.025em}}}

\def\RR{{\rm I\kern-1.6pt {\rm R}}}
\def\NN{{\rm I\!N}}
\def\ZZ{{\rm Z}\kern-3.8pt {\rm Z} \kern2pt}
\def\IB{\relax{\rm I\kern-.18em B}}
\def\ID{\relax{\rm I\kern-.18em D}}
\def\II{\relax{\rm I\kern-.18em I}}
\def\IP{\relax{\rm I\kern-.18em P}}
\newcommand{\CS}{{\scriptstyle {\rm CS}}}
\newcommand{\CSs}{{\scriptscriptstyle {\rm CS}}}
\newcommand{\rc}{\nonumber\\}
\newcommand{\bear}{\begin{eqnarray}}
\newcommand{\eear}{\end{eqnarray}}
\newcommand{\W}{{\cal W}}
\newcommand{\F}{{\cal F}}
\newcommand{\x}{{\cal O}}
\newcommand{\LL}{{\cal L}}

\def\mani{{\cal M}}
\def\calo{{\cal O}}
\def\calb{{\cal B}}
\def\calw{{\cal W}}
\def\calz{{\cal Z}}
\def\cald{{\cal D}}
\def\calc{{\cal C}}
\def\to{\rightarrow}
\def\ele{{\hbox{\sevenrm L}}}
\def\ere{{\hbox{\sevenrm R}}}
\def\zb{{\bar z}}
\def\wb{{\bar w}}
\def\nodiv{\mid{\hbox{\hskip-7.8pt/}}}
\def\menos{\hbox{\hskip-2.9pt}}
\def\dr{\dot R_}
\def\drr{\dot r_}
\def\ds{\dot s_}
\def\da{\dot A_}
\def\dga{\dot \gamma_}
\def\ga{\gamma_}
\def\dal{\dot\alpha_}
\def\al{\alpha_}
\def\cl{{closed}}
\def\cls{{closing}}
\def\vev{vacuum expectation value}
\def\tr{{\rm Tr}}
\def\to{\rightarrow}
\def\too{\longrightarrow}

%%%%%%%% ALBERTO's Macros %%%%%%%%%%%%%%%%
%%%%%%%%%%%%%%%%%%%%%%%%%%%%%%%%%%%%%%%%%%
%String variables (LaTeX 2e)
\newcommand{\gs}{\ensuremath{g_s}} % String coupling constant
\newcommand{\ap}{\ensuremath{\alpha'}} % Inverse string tension
\newcommand{\ls}{\ensuremath{l_s}} % String length
\newcommand{\ms}{\ensuremath{m_s}} % String scale
\newcommand{\lP}{\ensuremath{l_P}} % Planck length
\newcommand{\mP}{\ensuremath{m_P}} % Planck mass

%Useful definitions
\def\eps{\epsilon}
\def\del{\nabla}
\def\grad{\nabla}
\def\curl{\nabla\times}
\def\div{\nabla\cdot}
\def\p{\partial}
\def\pbar{\bar{\partial}}
\def\zbar{\bar{z}}
\def\para{{\scriptscriptstyle ||}}
\def\expec#1{\langle #1 \rangle}
\def\bra#1{| #1 \rangle}
\def\ket#1{\langle  #1 |}
\def\cbra#1{|\left. #1 \right)}
\newcommand{\Rea}{{\mathrm{Re}}}
\newcommand{\Ima}{{\mathrm{Im}}}
\newcommand{\cL}{{\mathcal{L}}}
\newcommand{\cZ}{{\mathcal{Z}}}
\newcommand{\bS}{{\mathbf{S}}}
\newcommand{\bT}{{\mathbf{T}}}
\newcommand{\bR}{{\mathbf{R}}}
\newcommand{\bZ}{{\mathbf{Z}}}
\newcommand{\cM}{{\mathcal{M}}}
\newcommand{\cG}{{\mathcal{G}}}

%Paper-specific macros
\newcommand{\tih}{\ensuremath{\tilde{h}}}
\newcommand{\Pt}{\ensuremath{P^{r}_{\theta}}}
\newcommand{\Px}{\ensuremath{P^{r}_{x}}}
\newcommand{\pix}{\ensuremath{\pi_{x}}}
\newcommand{\piphi}{\ensuremath{\pi_{\phi}}}
\newcommand{\pit}{\ensuremath{\pi_{\theta}}}

%
% FONTS

%\newfont{\headfont}{cmbx10 scaled 1440}
\newfont{\namefont}{cmr10}
%\newfont{\initialfont}{cmr10 scaled 1200}
\newfont{\addfont}{cmti7 scaled 1440}
\newfont{\boldmathfont}{cmbx10}
%\newfont{\figfont}{cmr7 scaled 1200}
\newfont{\headfontb}{cmbx10 scaled 1728}
\renewcommand{\theequation}{{\rm\thesection.\arabic{equation}}}
\begin{titlepage}

\begin{center} \Large \bf Drag Force in a Charged $\cN=4$ SYM Plasma

\end{center}

\vskip 0.3truein
\begin{center}
Elena C\'aceres$^{\dagger}$\footnote{elenac@ucol.mx} and
Alberto G\"{u}ijosa$^{*}$
\footnote{alberto@nucleares.unam.mx}
\vspace{0.3in}

${}^{\dagger}$  Facultad de Ciencias,\\ Universidad de Colima,\\
Bernal D\'{\i}az del Castillo 340, Colima, Colima, M\'exico\\
% {\small and}\\
% Theory Group, Department of Physics, University of Texas at Austin\\
% Austin, TX 78712, USA\\
 \vspace{0.3in}
$^{*}$ Departamento de F\'{\i}sica de Altas Energ\'{\i}as,
\\Instituto de Ciencias Nucleares, \\ Universidad Aut\'onoma de
M\'exico,\\ Apdo. Postal 70-543, M\'exico D.F.04510, M\'exico

\vspace{0.3in}

\end{center}
\vskip 1truein

{\bf ABSTRACT:} Following recent developments, we employ the AdS/CFT correspondence to
determine the drag force exerted on an external quark that moves through an $\cN=4$
super-Yang-Mills plasma with a non-zero R-charge density (or, equivalently, a non-zero
chemical potential). We find that the drag force is larger than in the case where the plasma
is neutral, but the dependence on the charge is non-monotonic.

\vskip2.6truecm \vspace{0.3in} \leftline{UTTG-07-06 } \vspace{0.2in}
\leftline{ICN-UNAM-06/05G} \vspace{0.2in}

\smallskip
\end{titlepage}
\setcounter{footnote}{0}

%--------+---------+--s-------+---------+---------+---------+---------+
%Body
\section{Introduction and Summary}

Central collisions of gold nuclei at RHIC are believed to produce the long-sought quark
gluon plasma (QGP). RHIC experiments have found evidence of strong collective behavior,  and
the consensus is that we are dealing with a strongly-interacting liquid that can be modelled
by hydrodynamics (see, e.g., \cite{Shuryak:2004cy} for a recent review). The hydrodynamic
regime is completely characterized by transport coefficients (shear and bulk viscosity,
etc.). RHIC data suggests that the QGP viscosity should be fairly small--- the ratio of
shear viscosity to entropy density has been estimated \cite{teaney} to be $\eta/s \sim
0.1-0.2$ , although uncertainties in this value are still quite large. Unfortunately, a
calculation of the hydrodynamic coefficients from first principles, {\it i.e}, from the
underlying microscopic theory, is currently out of our reach in the strong-coupling regime.

The gauge/gravity or AdS/CFT correspondence \cite{malda,gkpw}  has
been
 used to investigate  observables in various interesting strongly-coupled
 gauge theories  \cite{Csaki:1998qr}
 where perturbation
theory is not applicable.  Policastro, Son and Starinets pioneered the study of hydrodynamic
gauge theory properties  using AdS/CFT \cite{Policastro:2001yc,Policastro:2002se}. In
\cite{Policastro:2001yc}, these authors computed the shear viscosity $\eta$ of
strongly-coupled $\cN$=4 super-Yang-Mills (SYM) theory in $3+1$ dimensions and at finite
temperature. They found that the ratio of shear viscosity to entropy density equals
$1/{4\pi}$, a result that was later argued to be universal, in the sense that it applies to
any gauge theory described by a supergravity dual, in the limit of large 't Hooft coupling
\cite{Buchel:2003tz}.\footnote{The leading order correction in inverse powers of the 't
Hooft coupling was determined in \cite{Buchel:2004di}.}  These results raised the
tantalizing possibility of using AdS/CFT to study the QGP.
 Transport coefficients of different thermal gauge theories have
 been calculated in \cite{Policastro:2002se}-\cite{Benincasa:2006ei}.
 Work attempting to narrow the gap between the gauge/gravity
 duality and RHIC may be found in \cite{Sin:2004yx,Shuryak:2005ia}.

RHIC data also  confirmed the existence of jet quenching in high-energy heavy ion collisions
\cite{Adcox:2001jp} and  showed that the  observed suppression of hadrons from fragmentation
of hard partons is due to their interaction with the dense medium \cite{Adler:2002xw}. In a
series of interesting and very recent papers, the AdS/CFT machinery has been brought to bear
on the phenomenon of jet quenching and the associated parton energy loss. The authors of
\cite{Liu:2006ug} suggested that the jet-quenching parameter $\hat{q}$ used as a measure of
energy loss in phenomenological models  \cite{baier,kw} could be defined non-perturbatively
in terms of a lightlike Wilson loop, and then employed the gauge/gravity duality to compute
this loop in strongly-coupled $\cN=4$ SYM. The calculation was extended to the non-conformal
case in \cite{buchel}. The works \cite{hkkky,gubser} studied the dynamics of moving strings
in the AdS$_5$-Schwarzschild$\times\bS^5$ background to determine the drag force that a hot
neutral $\N=4$ SYM plasma exerts on a moving quark. In \cite{ct}, through an analysis of
small string fluctuations, the authors determined the diffusion coefficient for the plasma,
finding a result that is in agreement with \cite{hkkky}.

In this note we extend the computation of \cite{hkkky,gubser} to
the near-horizon geometry associated with rotating near-extremal
D3-branes, which allows us to examine how the drag force changes
upon endowing the plasma with a non-zero charge (or, equivalently,
chemical potential) under a $U(1)$ subgroup of the $SU(4)$
R-symmetry group.
The fact that the QGP produced at RHIC is also
found to carry a small but non-zero charge density (associated to
baryon number) cannot but add interest to our calculation. But of
course, since the string theory dual of QCD is not known, one
cannot overemphasize that, for multiple reasons, comparison of
AdS/CFT results with real data is, in general, a risky business
(for discussion, see \cite{hkkky,gubser}).

In Section \ref{bgsec} we review the relevant properties of the rotating black three-brane
geometry, and discuss the embedding of the Nambu-Goto string in this background. Section
\ref{dragsec} contains the calculation of the drag force. The dependence of our general
result (\ref{dragforce}) on the plasma temperature $T$ and charge density $J$ is shown in
Fig.~1, obtained by solving the sixth-order equation (\ref{sixth}) numerically. For small
charge density, this equation can be solved perturbatively, and (\ref{dragforcenlo}) gives
the resulting drag force at next-to-leading order. Our main conclusion is that the force in
the charged plasma case is \emph{larger} than in the neutral case. Interestingly, we find
the same result for the force in two cases where the external quark is coupled in different
ways to the SYM scalar fields, despite the fact that in one (`polar') case the quark is
neutral under the global $U(1)_R$, while in the other (`equatorial') case it is charged.

While the first version of this paper was in preparation an overlapping preprint
\cite{herzog} appeared. We will discuss its relation to our work at the end of Section
\ref{dragsec}.

\section{String in a Rotating D3-brane Background}
\label{bgsec}

The AdS/CFT correspondence \cite{malda,gkpw} equates the physics
of an external quark in a finite-temperature \emph{neutral}
$\cN=4$ SYM plasma to that of a fundamental string that lives on
the near-horizon geometry of a stack of static near-extremal
D3-branes \cite{wittenthermal}. This connection was employed in
\cite{hkkky,gubser} to determine the drag force that the neutral
plasma exerts on the quark.  Our aim in this paper is to consider
instead a plasma that is \emph{charged} under the $SU(4)$
R-symmetry, and so we must analyze a string that ploughs through a
rotating D3-brane background. For simplicity we will restrict
attention to the case where only one of the three $SU(4)$ angular
momenta is non-zero. The corresponding solution was obtained in
\cite{russo} (see also \cite{cveticgubser});\footnote{The metric
for the case with all three angular momenta turned on can be found
in \cite{klt}.} we will follow here the conventions of
\cite{gubserspinning}. In the near-horizon limit, the metric is
\begin{eqnarray} \label{metric}
ds^2&=&{1\over\sqrt{H}}(-hdt^2+d\vec{x}^2)+\sqrt{H}
\left[{dr^2\over\tilde{h}} -{2l r_0^2 \over R^2}\sin^2\theta
dtd\phi\right.
\\
{}&{}&\qquad\qquad\qquad\qquad\qquad\qquad\left.+r^2(\Delta
d\theta^2 +\tilde{\Delta}\sin^2\theta d\phi^2 + \cos^2\theta
d\Omega_3^2)\right], \nonumber
\end{eqnarray}
where
\begin{eqnarray}
H&=&\frac{R^4}{r^4\Delta}~,\nonumber\\
\Delta&=&1+\frac{l^2\cos^2\theta}{r^2}~,\nonumber\\
\tilde{\Delta}&=&1+{l^2\over r^2}~, \nonumber\\
h&=&1-{r_0^4\over r^4\Delta}~, \nonumber\\
\tilde{h}&=&{1\over\Delta}\left(1+{l^2\over r^2}-{r_0^4\over
r^4}\right)~,\nonumber\\
R^4&=& 4\pi N\gs\ls^4~. \label{L}
\end{eqnarray}
This geometry has an event horizon at the positive root of
$\tilde{h}(r_H)=0$,
\begin{equation} \label{rH}
r_H^2={1\over 2}\left(\sqrt{l^4+4r_0^4}-l^2\right)~,
\end{equation}
and an associated Hawking temperature, angular momentum density, and angular velocity at the
horizon
\begin{equation} \label{Tj}
T={r_H\over {2\pi R^2 r_{0}^2}}\sqrt{l^4+4r_0^4}~,\qquad J={l r_0^2 R^2\over 64
\pi^4\gs^2\ls^8}~,\qquad \Omega={l r_H^2\over r_0^2 R^2}~,
\end{equation}
which  translate respectively into the temperature, R-charge density and R-charge chemical
potential in the gauge theory. For $J\neq 0$, there is an ergosphere between $r=r_H$ and the
stationary limit surface $r=r_s(\theta)$ defined by $h(r_s,\theta)=0$.

An external quark (a pointlike source in the fundamental representation of the $SU(N)$ gauge
group) that moves at constant velocity in the $x^1\equiv x$ direction and carries a charge
under the same global $U(1)_R\subset SU(4)_R$ as the plasma corresponds to a string whose
embedding in the geometry (\ref{metric}) and in the static gauge
 $\sigma=r$, $\tau=t$ is of the general stationary form
\begin{equation} \label{embedding}
X^1(r,t)=vt+\xi(r), \qquad \theta(r,t)=\nu t+\zeta(r), \qquad
\phi(r,t)=\omega t+\chi(r),
\end{equation}
with all other fields vanishing. As usual, the behavior at infinity describes the gauge
theory in the extreme UV, where the pointlike quark is inserted, so the string must have a
single endpoint on the boundary. The corresponding boundary conditions\footnote{Notice that
an appropriate choice of origin allows us to set $\xi_{\infty}=0=\chi_{\infty}$ without loss
of generality.}
\begin{equation} \label{bc}
X(r,t)\to X_{\infty}(t)\equiv vt, \quad
\theta(r,t)\to\theta_{\infty}(t)\equiv\nu t+\zeta_{\infty}, \quad
\phi(r,t)\to\phi_{\infty}(t)\equiv\omega t \quad\mbox{as}\quad
r\to\infty,\end{equation}
 specify both the trajectory of the
quark (which determines its coupling $A_{\mu}(X_{\infty}(t))\p_t X^{\mu}_{\infty}(t)$ to the
gluonic fields,) and its global $U(1)_R$ charge (which controls its coupling
$\Phi_i(X_{\infty}(t))\hat{n}^i_{\infty}(t)$ to the scalar fields, with
$\hat{n}^i_{\infty}(t)\in\bR^6$ the unit vector which points towards the north pole of the
$\bS^3\subset\bS^5$ lying at polar angle $\theta_{\infty}(t)$ and azimuthal angle
$\phi_{\infty}(t)$ \cite{maldawilson}). The tail of the string is associated instead with a
flux tube that could be mapped out with the methods of \cite{dkk,cg}.\footnote{After this
work had appeared as a preprint on the arXiv, the relevant calculations were carried out
both for the case of vanishing \cite{gubseretal} and non-vanishing \cite{chinos} R-charge
density.} In the present case the tube which codifies a fixed gauge- and scalar-field
pattern that follows the moving quark and becomes wider as one moves back along the $-x$
direction at a given time (or, equivalently, as time elapses at a fixed location in $x$)
\cite{gubser}.

The equations of motion for the string are obtained from the
Nambu-Goto Lagrangian
\begin{equation}\label{nglagrangian}
\cL\equiv-\sqrt{-g}=-\sqrt{-\det (\p_a X^{\mu}\p_b
X^{\nu}G_{\mu\nu})}~,
\end{equation}
where $G_{\mu\nu}$ ($\mu,\nu=0,\ldots,9$) is the spacetime metric (\ref{metric}) and
$g_{ab}$ the induced metric on the worldsheet. The stationary embedding (\ref{embedding}) is
such that $\p_t(\p\cL/\p\dot{X}^{\mu})=0~\forall\mu$, and as a result, the equations for the
cyclic variables $X(r,t)$ and $\phi(r,t)$ amount to the statement that the conjugate spatial
momentum densities
\begin{equation}\label{pis}
\pix\equiv\pi^r_x= {\p\cL\over\p X'}~,\qquad
\piphi\equiv\pi^r_{\phi}={\p\cL\over\p\phi'}
\end{equation}
are constant. Our main goal is to determine the value of $\pix$, which controls the
$x$-component of the force that each segment of the string exerts on its larger-$r$ neighbor
\cite{hkkky},
\begin{equation}\label{piforce}
F_x={1\over 2\pi\ap}\pix~.
\end{equation}
In the situation of interest the string trails behind its boundary endpoint (which is being
pulled in the $+x$ direction by an external agent), so we should have $F_x<0$ and therefore
$\pix<0$.

The equation of motion for $\theta(r,t)$, on the other hand, is rather complicated, due to
fact that $\cL$ has explicit $\theta$-dependence. In particular, and in contrast with the
non-rotating case, a constant-$\theta$ solution is possible for $l\neq 0$ only if
$\theta_{\infty}=0,\pi$ or $\theta_{\infty}=\pi/2$, just like in the $v=0$ case studied
previously in \cite{bs}. For simplicity, we will restrict attention to these cases, which
describe a string that lies respectively perpendicular and parallel to the rotation plane of
the black brane.

\begin{itemize}
\item In the `polar' case $\theta_{\infty}=0$ or $\pi$, the string points at constant angles
$\theta(r,t)=0,\pi$ and $\phi(r,t)=0$, and reaches the horizon at the pole, where the
ergosphere has vanishing width ($\tilde{h}=h$, so the stationary limit surface coincides
with the horizon). Starting from the definition of $\pix$ in (\ref{pis}), it is then easy to
infer that
\begin{equation}\label{primespolar}
X'=-\pix\sqrt{G_{rr}\left(-G_{tt}/G_{xx}-v^2\right)\over -G_{tt}(-G_{tt}G_{xx}-\pix^2)}~,
\end{equation}
where, in accord with the previous discussion, the overall sign has been chosen in such a
way that $\pix<0$ implies $X'>0$, meaning that the string trails behind its boundary
endpoint.

\item In the `equatorial' case  $\theta_{\infty}=\pi/2$, the string is at constant polar
angle $\theta(r,t)=\pi/2$ but can have a nontrivial azimuthal profile $\phi(r,t)=\omega
t+\chi(r)$. One can again invert (\ref{pis}) to obtain $X'$ and $\phi'$ in terms of
$\pix,\piphi,v,\omega$. To simplify the algebra, we will henceforth concentrate on the case
$\omega=0$, where the string does not rotate. This choice is clearly valid from the point of
view of the gauge theory, where it amounts to specifying a time-independent coupling between
the external quark and the scalar fields. On the other hand, it might seem counterintuitive
from the gravity perspective, because we expect the string to penetrate beyond the
ergosphere, where a point particle could not avoid rotating. Nevertheless, we will see in
the next section that it is possible to find physical configurations where the string does
precisely that. Setting $\omega=0$, then, we find that the definitions (\ref{pis}) imply
\begin{eqnarray} \label{primesequatorial}
X'&=&\sqrt{G_{rr}\over G_{xx}(G_{t\phi}^2-G_{\phi\phi}G_{tt}) D}\left[\piphi
G_{t\phi}G_{xx}v+\pix\left\{-G_{t\phi}^2+G_{\phi\phi}
(G_{tt}+G_{xx}v^2)\right\}\right]~,\nonumber\\
\phi'&=&\sqrt{G_{rr}G_{xx}\over (G_{t\phi}^2-G_{\phi\phi}G_{tt}) D}\left[\piphi G_{tt}+\pix
G_{t\phi}v\right]~,\\
 D&\equiv & 2\piphi\pix
G_{t\phi}G_{xx}v+\pix^2\{-G_{t\phi}^2+G_{\phi\phi}
(G_{tt}+G_{xx}v^2)\}\nonumber\\
{}&{}&+G_{xx}\{\piphi^2
G_{tt}-(G_{t\phi}^2-G_{\phi\phi}G_{tt})(G_{tt}+G_{xx}v^2)\}~,\nonumber
\end{eqnarray}
where again the overall signs were chosen such that $\pix<0$ for the desired trailing
string.

\end{itemize}

\section{Drag Force in a Charged Plasma}
\label{dragsec}

We are now ready to compute the drag force exerted by the charged plasma on the quark,
following \cite{hkkky,gubser}. The first step is to demand that the solutions  to
(\ref{primespolar}) and (\ref{primesequatorial}) be well-defined over the entire region
$r_H< r<\infty$. This condition is nontrivial because the solutions involve square roots of
quantities that can become negative. As we will now learn, insisting that this does not
happen will fix the values of $\pix,\piphi$. As in \cite{hkkky,gubser}, in the analysis to
follow a special role will be played by the  velocity-dependent radius $r_v(\theta)$ that is
the largest root of the equation $h(r,\theta)-v^2=0$,
\begin{equation}\label{rvgral}
r_v(\theta)^2={1\over 2}\left(\sqrt{l^4\cos^4\theta +{4r_0^4\over
1-v^2}}-l^2\cos^2\theta\right)~.
\end{equation}

\begin{itemize}

\item In the polar case $X'$ is given by equation (\ref{primespolar}). The numerator
involves the factor $-G_{tt}/G_{xx}-v^2=h-v^2$, which according to (\ref{rvgral}) changes
sign at
\begin{equation}\label{rvpolar}
r_v(0)=\sqrt{{1\over 2}\left(\sqrt{l^4+{4r_0^4\over
1-v^2}}-l^2\right)}=r_v(\pi)~.
\end{equation}
The only way we can prevent $X'$ from becoming imaginary for
$r<r_v(0)$ is by choosing a value of $\pix$ such that the
denominator also changes sign at $r_v(0)$, i.e.
\begin{equation}\label{pispolar}
\pix=-\sqrt{h(r_v)\over H(r_v)} = -{r_0^2\over R^2}{v\over
\sqrt{1-v^2}}~.
\end{equation}
The string profile can then be determined by integrating the equation obtained by plugging
(\ref{pispolar}) into (\ref{primespolar}),
\begin{equation}\label{solpolar}
X'=v{r_0^2\over R^2}{H\over h}={vr_0^2 R^2 \over r^4+l^2 r^2-r_0^4}~.
\end{equation}

\item In the equatorial case $X'$ and $\phi'$ are given by
(\ref{primesequatorial}). Using the explicit metric components
(\ref{metric}), it is easy to convince oneself that, out of the
various factors inside the square roots, only the function
\begin{eqnarray}\label{delta}
 D(r)&=&\left[r_0^4 \pix^2 +{r_0^8\over R^4}\right]r^{-2} +
\left[{r_0^4\over R^4}\piphi^2-2v l{r_0^2\over
R^2}\piphi\pix- (1-v^2)l^2\pix^2 -l^2 {r_0^4\over R^4}\right]r^0\\
{}&{}&+\left[-(1-v^2)\pix^2-(2-v^2){r_0^4\over R^4}\right]r^2 +
\left[(1-v^2){l^2\over R^4}-{\piphi^2\over R^4}\right]r^4
+\left[{1-v^2}\over R^4\right]r^6\nonumber
\end{eqnarray}
runs the risk of becoming negative. This function clearly approaches $+\infty$ both at $r\to
0$ and $r\to\infty$, so it must have at least one minimum at some intermediate point. Since
$r^2 D(r)$ is quartic in the variable $r^2$, it could in general have as many as four
distinct roots at positive values of $r$. Numerical calculation shows that, for generic
values of $\pix,\piphi$, the function $ D(r)$ has exactly two positive roots, $r_1,r_2>r_H$,
and a single minimum in between. This implies that $ D(r)<0$ (and the solution
(\ref{primesequatorial}) is ill-defined) in the intermediate range $r_1<r<r_2$, an undesired
feature that can only be avoided by choosing $\pix,\piphi$ in such a way that the $ D(r)$
curve has its minimum at $ D(r_1=r_2)=0$. Analytically, this corresponds to demanding that
the $r^0$ and $r^4$ terms in (\ref{delta}) vanish, implying that
\begin{eqnarray}\label{pisequatorial}
\piphi&=&l\sqrt{1-v^2}~,\nonumber\\
\pix&=&-{r_0^2\over R^2}{v\over \sqrt{1-v^2}}~,
\end{eqnarray}
after which $ D$ is found to take the form
\begin{equation}\label{deltasimp}
 D(r)={1-v^2\over R^4}\left[r^4-{r_0^4\over 1-v^2}\right]^2~.
\end{equation}
This function is manifestly non-negative, and evidently has a minimum and a double root at
$r_0/(1-v^2)^{1/4}$. Notice that this is none other than the critical radius $r_v(\pi/2)$
defined in (\ref{rvgral}).

The string profile in this case follows from integration of the equations of motion
(\ref{primesequatorial}), which simplify drastically after use of (\ref{pisequatorial}):
\begin{eqnarray}\label{solequatorial}
X'&=&{vr_0^2 R^2 \over r^4+l^2 r^2-r_0^4}~,\\
\phi'&=&-{lr^2 \over r^4+l^2 r^2-r_0^4}~. \nonumber
\end{eqnarray}
These expressions are clearly well-defined over the entire range $r_H<r<\infty$. One might
have also worried that the string worldsheet could become spacelike at some point inside the
ergosphere, but in fact (\ref{solequatorial}) leads to $\sqrt{-g}=\sqrt{1-v^2}$, which is
manifestly real. So as promised, we have been able to find a physical configuration where
the string penetrates into the ergosphere and yet does \emph{not} rotate.\footnote{In the
process, we have also learned that the solution ran the risk of being ill-defined not just
inside the ergosphere, but in the region below the critical radius $r_v(\theta)$ defined in
(\ref{rvgral}), which coincides with the stationary limit surface $r_s(\theta)$ only for
$v=0$. } What has happened is that the string has managed to adopt an azimuthal profile
$\phi(r)$ that leads to a precise balance between the string tensile force and the inertial
drag due to the rotating geometry. Notice that, according to (\ref{pisequatorial}), this
requires an external agent to pull on the boundary endpoint of the string, exerting the
non-zero force $\piphi$. Notice also that the sign of $\phi'$ in (\ref{solequatorial})
correctly reflects the fact that, as $r$ increases, the string should wind in the direction
opposite to the sense of rotation.
\end{itemize}

The second step is to use (\ref{pispolar}) or (\ref{pisequatorial}) in (\ref{piforce}), to
read off the force with which the string pulls back on its boundary endpoint, which by the
AdS/CFT correspondence should be identified with the drag force that the charged plasma
exerts on the moving quark. Surprisingly, the result is the same in the polar and equatorial
cases,
\begin{equation} \label{dragforce}
F_x={dp_x\over dt}=-{{r_0^2/R^2}\over 2\pi\ls^2}{v\over\sqrt{1-v^2}}~.
\end{equation}
We learn then that the value of the $U(1)_R$ charge of the external quark affects the
gluonic and scalar field distributions set up by the quark (which, as explained in Section
\ref{bgsec}, are encoded in the shape of the string tail (\ref{solpolar}) or
(\ref{solequatorial})), but does not alter the total drag force experienced by the quark.
Notice that the end result (\ref{dragforce}) is identical to the one given in equation (12)
of \cite{gubser}, except for the fact that $r_H$ has been replaced here by $r_0$. The two
radii indeed agree for the non-rotating case, but are related through (\ref{rH}) in the
general case, which shows that the drag force \emph{does} have an interesting dependence on
the R-charge density of the plasma.

The third and final step is to rewrite (\ref{dragforce}) in terms of gauge theory
parameters. Using (\ref{rH}) and (\ref{Tj}) one finds that $\rho\equiv r_0^4/16\pi^4 R^8
T^4$ satisfies the sixth-order equation
\begin{equation} \label{sixth}
16\rho^6-\rho^5-4c^2 \rho^4 + 8c^4\rho^3-c^6\rho+c^8=0~,
\end{equation}
where we have defined a dimensionless charge parameter
\begin{equation}\label{c}
c\equiv {J\over 2\pi N^2 T^3}~.
\end{equation}

In the neutral case $c=0$, equation (\ref{sixth}) implies that $\rho=1/16$, from which one
immediately recovers the well-known relation $r_0=\pi R^2 T$. For small charge densities one
can obtain $r_0$ by solving (\ref{sixth}) in an expansion in powers of $c$. The result at
next-to-leading order is
\begin{equation} \label{r0nlo}
r_0=\pi R^2 T \left[1+{4 J^2\over \pi^2 N^4 T^6} + \cO(J^4/N^8 T^{12})\right]~,
\end{equation}
which together with (\ref{L}), (\ref{dragforce}), $p_x=mv/\sqrt{1-v^2}$ and $g_{YM}^2\equiv
4\pi\gs$ leads to the final result
\begin{equation} \label{dragforcenlo}
{dp_x \over dt}=-{\pi\over 2}{\sqrt{g_{YM}^2 N}T^2}{p_x\over m}\left[1+{8 J^2\over \pi^2 N^4
T^6} + \cO(J^4/N^8 T^{12})\right]~.
\end{equation}
{}From this we can read off the exponential relaxation time
\begin{equation} \label{decayconstant}
\tau_0={2\over\pi\sqrt{g_{YM}^2 N}}{m\over T^2}\left[1-{8 J^2\over \pi^2 N^4 T^6} +
\cO(J^4/N^8 T^{12})\right]~.
\end{equation}
We learn here that turning on a nonzero R-charge density for the
plasma \emph{increases} the drag force exerted on the heavy quark
(or equivalently, decreases its relaxation time). Notice that the
$J$-dependent terms involve inverse powers of $N$, as has been
found when computing other properties of the charged plasma (e.g.,
its entropy density \cite{gubserspinning}).

\begin{figure}[htb]
\begin{center}
\setlength{\unitlength}{1cm}
\includegraphics[width=9cm,height=6cm]{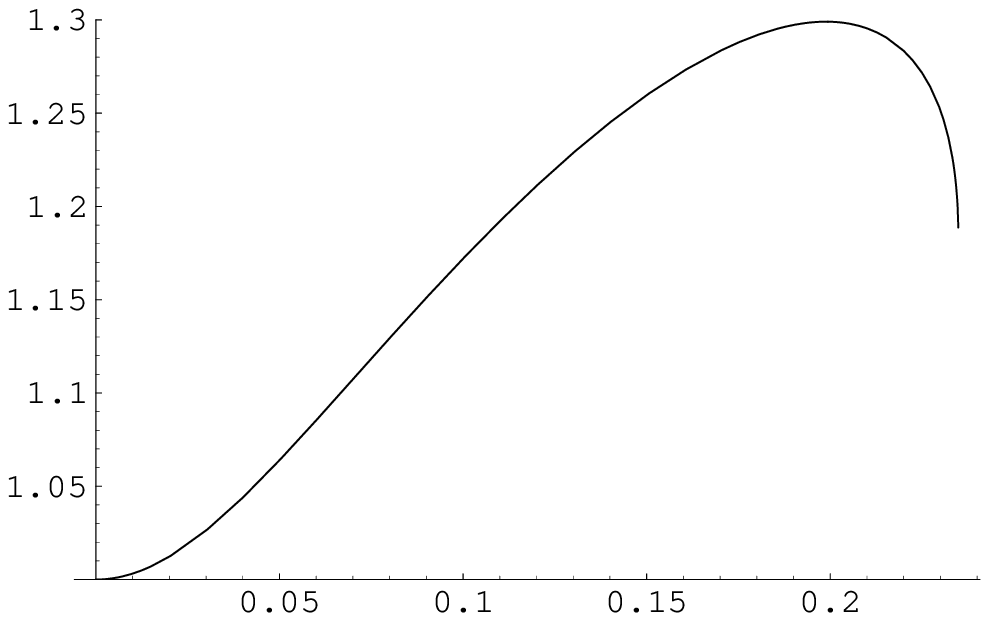}
 \begin{picture}(0,0)
   \put(0.2,0.3){$c$}
   \put(-11.2,5.6){$F_x(c)/F_x(0)$}
 \end{picture}
\caption{Drag force as a function of the dimensionless charge
parameter $c=J/ 2\pi N^2 T^3$.} \label{ratiofig}
\end{center}
\end{figure}

The behavior of the drag force for larger charge densities can be
determined by solving (\ref{sixth}) numerically. The result is
shown in Fig.~\ref{ratiofig},\footnote{The original preprint that
we posted to the arXiv gave the arbitrary-charge result only
implicitly, through (\ref{sixth}). We have added the plot to this
later version in order to facilitate comparison with
\cite{herzog,as}.} which displays the ratio between the force in
the charged case and that in the neutral case. Real solutions
exist only in the finite range $0\le c\le c_{max}\simeq 0.2349$.
The main feature is the maximum at $c_0\simeq 0.199$, where the
force ratio reaches a value $\simeq 1.299$, and beyond which it
decreases to $\simeq 1.184$ at $c_{max}$.

As explained in the Introduction, the authors of \cite{Liu:2006ug}
advocated the use of a lightlike Wilson loop to define a parameter
$\hat{q}$ meant to be an alternative characterization of energy
loss in a thermal plasma. The works
\cite{cg2,chinojapones,as,edelstein} followed this procedure to
determine $\hat{q}$ for the charged $\cN=4$ SYM plasma, obtaining
results that are rather similar to the drag force computed above.
The drag force curve given in Fig.~\ref{ratiofig} is in fact
nearly identical to the analogous $\hat{q}$ curve depicted (as a
function of $\xi\equiv 4\sqrt{2}c$) in Fig.~2 of \cite{as}.
(Particularly striking is the fact that the two ratios coincide at
$\xi_{max}=4\sqrt{2}c_{max}$.) This suggests that, at least in the
current setting, both quantities indeed capture essentially the
same physics.

It is also interesting to compare our results with those obtained in the closely related
work \cite{herzog}, which appeared while the first version of this paper was in preparation.
After carrying out a general drag force calculation applicable to any gauge theory whose
dual gravity description involves an asymptotically AdS$_{d+1}$ geometry, the author of
\cite{herzog} considers as a specific example the case of $d=4$ R-charged $\cN=4$ SYM
plasma, with all three of the independent charge parameters (corresponding to the
independent rotations within $SU(4)_R$) turned on. In the present paper we have examined
precisely this system, in the special subcase where two of these parameters are taken to
vanish.

There is, however, an important difference in our calculations: whereas the string studied
here moves in the ten-dimensional background (\ref{metric}), which is a solution of Type IIB
supergravity describing the near-horizon geometry of a stack of rotating D3-branes, the
string analyzed in \cite{herzog} lives  on the five-dimensional `STU' background, which
corresponds to a charged black hole solution of $\cN=8$ $D=5$ gauged supergravity
\cite{bcs}. The latter theory is believed to be a consistent truncation of the Kaluza-Klein
$\bS^5$ reduction of Type IIB supergravity, and in \cite{cveticgubser} it was shown that the
STU and spinning D3 backgrounds are indeed related in this manner: the background described
by equations (5.1)-(5.5) of \cite{herzog} is shown there to be a non-trivial truncation of
our (\ref{metric}), with the identifications $\kappa_1 = (l/r_H)^2$, $\kappa_2 = \kappa_3 =
0$. The drag force deduced from the STU background in this case is given in equation (5.12)
of \cite{herzog}, which is then to be compared with our spinning D3 result
(\ref{dragforce}). The two forces are clearly \emph{very} different: the result of
\cite{herzog} displays a complicated dependence on the velocity $v$ and the charge parameter
$\kappa=(l/r_H)^2$, which is completely unlike the simple dependence seen in
(\ref{dragforce}).

This discrepancy seems to suggest that: i) the putative uplift from five to ten dimensions
of the string configuration considered in \cite{herzog} would have a complicated (but
presumably still stationary, as in (\ref{embedding}),) radial dependence for both the polar
and the azimuthal angles, unlike the polar and equatorial profiles considered here;  ii)
string configurations more general than the ones considered here (e.g., a non-polar string
dual to a quark whose coupling with the SYM scalar fields is time-independent) could
potentially experience a drag force whose functional dependence on the relevant parameters
is much more complicated than the one seen in (\ref{dragforce}). It would be interesting to
explore this second point by searching for explicit configurations of this type in ten
dimensions. The first point is obscured by the non-trivial nature of the truncation that
connects the two backgrounds. In particular, given that the string of \cite{herzog} does not
couple to the five-dimensional gauge field $A_t\,^{\phi}=-l r_0^2/R^2 r^2\tilde{\Delta}$,
which descends from the ten-dimensional metric component $G_t\,^{\phi}$ \cite{cveticgubser},
it would seem like the angular dependence of its uplift should be special enough to somehow
decouple the angular problem from the five-dimensional geometry, which is all that is
captured by the Nambu-Goto action used in \cite{herzog}. Our polar string seems to do
precisely that, so it is not clear why this configuration appears not to be accessible from
the five-dimensional perspective.

A second possibility is that the string of \cite{herzog} is in some sense smeared over the
$\bS^5$. This would imply that its dual quark is qualitatively different from the one
considered here, because its coupling to the SYM scalar fields would be somehow averaged
over. A third possibility that should be kept in mind is that, even though the
categorization of $\N=8$ $D=5$ supergravity as a consistent truncation of Type IIB
\emph{supergravity} would guarantee that any solution of the former can be uplifted to a
solution of the latter, it could turn out not to be possible to capture the dynamics of a
\emph{string} that lives in the uplifted background in terms of a Nambu-Goto action that
senses only the five-dimensional geometry.

The results obtained in \cite{hkkky,gubser,herzog} and the present paper refer to a colored
object, the external quark. For comparison, it might be worth computing the drag force on a
color-neutral object, such as a meson\footnote{After this work had appeared as a preprint on
the arXiv, the corresponding calculations were carried out in \cite{psz,lrw2,cgg}, and
related work was reported in \cite{elena,argyres,as2,gubseretal3}.}
\cite{reyee,maldawilson,wittenthermal} or a baryon \cite{imamura,cgs,cgst}. We hope to
return to this and related problems in future work.

\section{Acknowledgments}
It is a pleasure to thank Alejandro Ayala for useful discussions on RHIC physics, Hernando
Quevedo for a conversation on Kerr-AdS black holes, and David Vergara for pointing out some
valuable references. We are also grateful to Chris Herzog and our JHEP referee for very
useful observations. Elena C\'aceres thanks the Theory Group at the University of Texas at
Austin for hospitality during the completion of this work. Her research is supported in part
by the National Science Foundation under Grant No. PHY-0071512 and PHY-0455649 and by a
Ram\'on Alvarez Buylla grant 5783. Alberto G\"uijosa is grateful for the hospitality of the
Universidad de Colima, which allowed this work to be carried out. His research is supported
in part by Mexico's National Council for Science and Technology grants CONACyT 40754-F,
CONACyT SEP-2004-C01-47211 and CONACyT/NSF J200.315/2004, as well as by DGAPA-UNAM grant
IN104503-3.

\end{document}